\documentclass[aps,prl,10pt,twocolumn,floatfix,noshowpacs,preprintnumbers,superscriptaddress,nofootinbib]{revtex4-2}
\usepackage[utf8]{inputenc}
\usepackage{amsmath}
\usepackage{amssymb}
\usepackage[T1]{fontenc}
\usepackage{bbold}
\usepackage{enumerate}
\usepackage{tikz}
\usepackage[compat=1.1.0]{tikz-feynman}
\usepackage{feynmf}
\usepackage{slashed}
\usepackage{braket}
\usepackage{url}
\usepackage{multirow}
\usepackage{natbib}
\usepackage{graphicx}
\usepackage[pdftex,bookmarks,linktocpage,pdfpagelabels,plainpages=false,hyperfigures,linkcolor=blue,citecolor=blue]{hyperref}
\hypersetup{colorlinks=true}
\usepackage[capitalise]{cleveref}
\usepackage{isotope}
\usepackage{mathrsfs,amssymb}
\usepackage{cancel}
\usepackage[normalem]{ulem}
\usepackage{orcidlink}
\usepackage{xspace}

\newcommand{\src}{s}  
\newcommand{\SNnorm}{\Phi^\star/\Phi^\star_0}

\newcommand{\hatwcal}{\widehat{\mathbf{w}}_{\rm cal}}
\newcommand{\hatw}{\widehat{\mathbf{w}}}

\newcommand{\barwsn}{\mathbf{w}^\star_{\rm cal}}

\newcommand{\bfm}[1]{\mathbf{#1}}
\newcommand{\MARLEYvtwo}{\texttt{MARLEY} \!\texttt{v2}\xspace}

\definecolor{tc_color}{rgb}{1.0, 0.66, 0.07}

\begin{document}

\preprint{FERMILAB-PUB-26-0418-T}

\author{Ting Cheng}
\email{tcheng1@fnal.gov}
\affiliation{Theoretical Physics Department, Fermilab, P.O. Box 500, Batavia, IL 60510, USA}

\author{Matheus Hostert}
\email{matheus-hostert@uiowa.edu}
\affiliation{Department of Physics and Astronomy, University of Iowa, Iowa City, IA 52242, USA}

\author{Pedro~A.~N.~Machado}
\email{pmachado@fnal.gov}
\affiliation{Theoretical Physics Department, Fermilab, P.O. Box 500, Batavia, IL 60510, USA}

\author{Nityasa Mishra}
\email{nityasa_mishra@tamu.edu}
\affiliation{Mitchell Institute for Fundamental Physics and Astronomy, Department of Physics and Astronomy, Texas A\&M University, College Station, Texas 77843, USA}

\author{Adrian Thompson}
\email{a.thompson@northwestern.edu}
\affiliation{Department of Physics \& Astronomy, Northwestern University, 2145 Sheridan Road, Evanston, IL 60208, USA}

\title{Standard Candles for Supernova Neutrino Detection at DUNE}

\begin{abstract}
The Deep Underground Neutrino Experiment (DUNE) far detector is sensitive to $\mathcal{O}(10)$ MeV electron neutrinos through $\nu_e$ charged-current reaction with argon. 
This capability is a unique window into the $\nu_e$ component of a Galactic core-collapse supernova flux. 
Extracting the properties of the supernova spectrum is, however, limited by the poorly-known $\nu_e$-Ar cross section. 
We propose a data-driven strategy that leverages $^8$B solar neutrinos and muon-decay-at-rest neutrinos as standard candles for this process.
These calibration samples constrain both the low-energy and high-energy components of the cross section relevant for supernova detection.
Our method reduces the reliance on nuclear models, which can bias the extraction of the spectral parameters by as much as 300\%.
\end{abstract}

\maketitle

The first observation of supernova (SN) neutrinos from SN1987A opened numerous research directions~\cite{Scholberg:2012id}.
This is despite the very limited number of events, namely 8 events at IMB~\cite{Bionta:1987qt,IMB:1988suc}, 12 events at Kamiokande-II~\cite{Kamiokande-II:1987idp,Hirata:1988ad}, and 5 events at Baksan~\cite{Alekseev:1988gp}. 
While these events have taught us a lot, they still cannot make definitive statements about the physics of supernova explosions, the neutrino mass ordering, or potential new physics.\footnote{Indeed, it has recently been pointed out that a mild disagreement between the data and supernova models is already present in SN1987A data~\cite{Li:2023ulf}.}
With an average rate of one supernova in the Milky Way per $40 \pm 10$~years~\cite{Diehl:2006cf,Rozwadowska:2020nab,Tammann:1994ev}, the chance of catching the next galactic SN is $\mathcal{O}(20\%)$ per decade, and it will be detected in a much more prepared and diverse network of detectors~\cite{Adams:2013ana}.
Super-Kamiokande currently provides the leading sensitivity to galactic supernova fluxes~\cite{Super-Kamiokande:2022bwp}, while other experiments deliver complementary measurements, including SNO+~\cite{Caden:2024cgr,Wang:2022egn}, KamLAND~\cite{KamLAND:2015dbn}, IceCube~\cite{IceCube:2011cwc}, NOvA~\cite{NOvA:2020dll}, the LAr detectors in the SBN program, and dark matter direct detection experiments~\cite{Lang:2016zhv,XLZD:2024nsu,DarkSide20k:2020ymr}.
Many preparations are already underway at DUNE~\cite{DUNE:2020zfm,DUNE:2023rtr}, Hyper-Kamiokande~\cite{Hyper-Kamiokande:2021frf}, and JUNO~\cite{JUNO:2015zny}, where ${\sim}10^3$ to $10^5$ supernova neutrino events could be anticipated if one occurs within $10$~kpc~\cite{Mirizzi:2015eza}.

Because neutrinos and antineutrinos carry away most of the gravitational binding energy released in core collapse, their flavor-dependent spectra encode the collapse, accretion, shock evolution, and proto-neutron star cooling history~\cite{Janka:2006fh, Janka:2012wk, Mirizzi:2015eza}. 
These time slices of spectra therefore probe the explosion mechanism~\cite{Janka:2012wk, Tamborra:2013laa, Burrows:2020qrp}, 
dense-matter physics~\cite{Martinez-Pinedo:2012eaj, Roberts:2016rsf}, 
flavor evolution in extreme environments~\cite{Duan:2010bg, Tamborra:2020cul, Volpe:2023met}, 
and possible new physics~\cite{Raffelt:1996wa,Raffelt:2025wty}. 
DUNE is essential to this program because it supplies the high-statistics charged-current $\nu_e$ sample on argon, complementing the predominantly $\bar\nu_e$ information from the rest of the global network. 

The dominant SN signal channel at DUNE is neutrino capture on an argon nucleus, namely, the inclusive charged-current reaction
\begin{equation}
	\label{eq:interaction}	
 	\nu_e + \isotope[40]{Ar} \to e^- + X \, ,
\end{equation}
where $X$ denotes any hadronic final state, including the discrete spectrum of excitations of $\isotope[40]{K}$ at low neutrino energies and the continuum of states with one or more emitted nucleons above the corresponding energy thresholds for $E_\nu \sim 10$~MeV.
The associated cross section, $\sigma$, was until recently unmeasured: first evidence from $^8$B solar neutrinos in DEAP-3600~\cite{Adhikari:2026tcd} yields an energy-averaged cross section that is $\sim 2.4^{+1.3}_{-1.0}$ times larger than predicted in Ref.~\cite{Bhattacharya:2009zz}.

The discrete contribution to neutrino capture is determined by the Fermi and Gamow–Teller transition strengths, which are inferred from a combination of complementary probes, including $\isotope[40]{Ti}$ $\beta^+$ decay~\cite{Bhattacharya:1998hc, ICARUS:1998nzl}, (p,n) charge-exchange~\cite{Bhattacharya:2009zz} and photon scattering~\cite{Li:2006qs,Gayer:2019eed,Tornow:2022kmo} reactions on Ar, and nuclear structure data for $\isotope[40]{K}$~\cite{Chen:2017ngq}.
The continuum contribution was studied in detail in Ref.~\cite{Gardiner:2026psy}.
Existing predictions differ widely across the SN energy range~\cite{Raghavan:1986hv,Ormand:1994js,GilBotella:2003sz,Kolbe:2003ys,Samana:2008cr,Krmpotic:2009ad,Cheoun:2011zza,Suzuki:2012ds,Paar:2012dj,Barbero:2020bmg,Gardiner:2021qfr,Gardiner:2026psy}.
Therein lies the problem: an incorrect cross section model can strongly bias the inferred spectral parameters from DUNE data~\cite{DUNE:2023rtr}.

In this \emph{letter}, we propose a data-driven approach to overcome this problem. 
We show that the combination of the detection of solar $^8$B neutrinos at DUNE~\cite{Capozzi:2018dat, Meighen-Berger:2024xbx} with muon decay-at-rest ($\mu$DAR) at stopped-pion sources can provide percent-level precision calibration measurements for neutrino capture on argon at low and high energies, respectively.
The uncertainty on the $^8$B flux is at the $\sim 4\%$ level from measurements at SNO~\cite{SNO:2002tuh,SNO:2024vjl} and Super-Kamiokande~\cite{Super-Kamiokande:2023jbt}, and is likely to improve to $\sim 1\%$ with future measurements at Hyper-Kamiokande~\cite{Hyper-Kamiokande:2021frf}, JUNO~\cite{JUNO:2015zny}, and next-generation direct-detection experiments such as XLZD~\cite{XLZD:2024nsu}.
The flux shape is also precisely and robustly constrained by $\beta$-decay kinematics~\cite{Longfellow:2023hoj,Bhattacharya:2006ah,Kirsebom:2011zz,Roger:2012zz}.
The \(\nu_e\) flux from the Sun relevant for DUNE is \(\phi^\odot = \phi^\odot_0 P_{ee}\)~\cite{Parke:1986jy, Bahcall:2004pz}.
Since \({}^8\)B neutrinos undergo adiabatic MSW evolution in the Sun and exit dominantly as the \(\nu_2\) eigenstate, \(P_{ee} \simeq |U_{e2}|^2 \simeq \sin^2\theta_{12}\), which will be precisely determined by JUNO~\cite{JUNO:2015zny}.
Note that \emph{hep} neutrinos are subdominant.

For $\mu$DAR, the primary calibration sample will come from the COHERENT program at the Oak Ridge Spallation Neutron Source (SNS) with the measurement of the capture cross section at the near-term COH-Ar-750 detector~\cite{COHERENT:2022nrm,COHERENT:2026ewu}.
The SNS neutrino flux can be constrained to the $2$--$3\%$ level thanks to the COHERENT D$_2$O flux monitor~\cite{COHERENT:2021xhx,COHERENT:2022nrm, COHERENT:2026ewu} and the $\mu$DAR flux shape follows the well-known three-body kinematics of muon decay.\footnote{Other argon-based detectors such as CCM~\cite{ChavezEstrada:2025ccm} or MicroBooNE, SBND, and ICARUS can in principle also measure $\nu_e$ capture on Argon from $\mu$DAR albeit with limited statistics~\cite{BenevidesRodrigues:2022wxz}.} 

The SN $\nu_e$ spectrum has mean energies of order $10$~MeV, spanning roughly $5$--$20$~MeV across emission phases, progenitor masses, and flavor-conversion scenarios~\cite{Fischer:2009af,Serpico:2011ir,Fiorillo:2023frv,Raffelt:2025wty}.
This range is bracketed by the two calibration sources: solar $^8$B neutrinos from below ($\langle E_\nu\rangle \simeq 7$~MeV) and $\mu$DAR $\nu_e$ from above ($\langle E_\nu\rangle \simeq 32$~MeV).
While solar neutrinos excite a fraction of the nuclear transitions in $\nu_e$-Ar capture that are relevant for SN, $\mu$DAR neutrinos excite all of them. 
Combined, they offer greater calibration power and, therefore, serve as \emph{standard candles}.

We parameterize the cross section in a deliberately over-complete, model-independent basis: arbitrary weights on a dense grid of nuclear transition energies.
While unphysical, our parameterization accommodates arbitrary scattering  \emph{modes} while maintaining the leptonic kinematics fixed.
No single data set can determine all of these weights, but none needs to.
A principal-component analysis of the predicted solar and $\mu$DAR spectra identifies the few combinations of weights that the calibration data constrain at the few-percent level.
A Fisher-information analysis then identifies the few modes that contribute to the bulk of the SN event rate spectrum.

\textit{The cross section.}---Following Refs.~\cite{Bhattacharya:1998hc, Gardiner:2021qfr, Gardiner:2026psy}, we write the cross section for neutrino capture $\nu_e + \isotope[40]{Ar} \to e^- + X$ as 
\begin{equation}
  \frac{d\sigma}{d T_e}=\sigma_0\,\mathcal{S}(T_e,E_\nu) \, ,  
\end{equation}
where $\sigma_0=8\times10^{-45}~{\rm cm}^2$ sets the normalization and 
\begin{equation}
\label{eq:xsec}
\mathcal{S}(T_e,E_\nu)
= \sum_k w_k\left(\frac{E_e p_e}{{1~\rm MeV}^2}\right) \delta(E_e - E_\nu + \Delta E_k) \, ,
\end{equation}
for incoming neutrino energy $E_\nu$, energy differences $\Delta E_k$, and electron kinetic energy $T_e$ and momentum $p_e$.
The leptonic phase-space factor $E_e p_e$ is kept explicit.
The $(E_\nu,E_e)$ dependence assumed in \cref{eq:xsec} may not accurately describe the underlying physics, especially at high energies, but we have checked that our simple choice leads to a small bias in the extraction of SN parameters.
More general parameterizations of the differential cross section can reduce the bias further.

The different weights $w_k$ are determined by both discrete and continuous excitations of argon.
For each transition, such as $\nu_e\,^{40}{\rm Ar}\to e^- n\,^{39}{\rm K}$, a different weight could in principle be calculated by ab-initio nuclear models or determined from data.
However, in our analysis, we adopt a large number of free weights $w_k$ such that the cross section has enough freedom to fit the data.
By taking a dense grid of $\Delta E_k$, this parametrization becomes over-complete and can cover both discrete and continuous spectra.
We distribute \(\Delta E_k\) uniformly between the reaction threshold $\Delta E_{\rm min}\approx 1.5~{\rm MeV}$ and \(45~{\rm MeV}\) using 200 points; this range covers the lowest transition energies and the high-\(E_\nu\) tail of $\mu$DAR and SN fluxes.

\textit{Count Spectra}---For each source $\src \in \{\odot,\star,\ominus\}$, corresponding to solar, supernova, and muon $\mu$DAR measurements, the predicted event count in reconstructed electron energy bin $i$ is
\begin{equation}
\label{eq:rate}
    n_i^\src =
    \sigma_0\,\tau^\src
    \int dE_\nu\, dT_e\,
    \frac{d\phi^\src}{dE_\nu}\,
    \mathcal{S}(T_e,E_\nu)\,
    \epsilon_i(T_e),
\end{equation}
where \(\tau^\src\) is the number of argon targets times the exposure time, \(d\phi^\src/dE_\nu\) is the  differential neutrino flux, and \(\epsilon_i(T_e)\) is the fraction of events with true electron kinetic energy \(T_e\) that migrate into reconstructed bin \(i\). 
The energy resolution at DUNE (COH-Ar-750) is modeled as a Gaussian smearing with fractional resolution $11\%/\sqrt{T_e/{\rm MeV}}$~\cite{DUNE:2020zfm} $(5\%/\sqrt{T_e/{\rm MeV}}$~\cite{COHERENT:2026ewu}$)$.
For \(\mathcal{S}(T_e,E_\nu)\), we use \MARLEYvtwo\cite{Gardiner:2021qfr,Gardiner:2026psy} to generate mock data and the phenomenological parameterization in \cref{eq:xsec} to carry out our analysis.
Our calibration procedure is independent of the details of the cross section model used to generate mock data.
Substituting \cref{eq:xsec} into \cref{eq:rate} yields the linear form
\begin{equation}
\label{eq:linear}
    \mathbf{n}^\src=R^\src \mathbf{w} ,
\end{equation}
where the rectangular \emph{response matrix} \(R^\src\) maps the vector of cross section weights \(\mathbf{w}\) (chosen to be 200-dimensional) to the vector of events in reconstructed electron energy bins.
The superscript $s$ denotes the source.

\begin{table}[t]
\begin{ruledtabular}
\begin{tabular}{clccc}
&Measurement & Flux [$\nu_e/{\rm cm}^2/{\rm s}$] & Exposure & $N_{\rm evts}$ \\
\hline
\(\odot\)& Solar & \(1.6\times10^6\) & \(10~{\rm kt\cdot yr}\) & \(1.0\times 10^4\) \\
\(\ominus\)& \(\mu\) decay-at-rest & \(7.4\times10^6\) & varies & \(3.4\times10^2\) \\
\(\star\)& Supernova & varies & \(40~{\rm kt}\) & \(1.0\times 10^3\) \\
\end{tabular}
\end{ruledtabular}
\caption{Benchmark parameters for each measurement.
The number of measured events expected is fixed and shown as $N_{\rm evts}$; for the supernova, this is the time-integrated total over all emission phases.}
\label{tab:benchmarks}
\end{table}

\Cref{tab:benchmarks} summarizes the benchmark fluxes, exposures, and event counts for the sources in question.
For solar neutrinos at DUNE, we take a $10~{\rm kt\cdot yr}$ exposure representative of an early measurement at the far detector, before the full $40~{\rm kt}$ fiducial mass is available.
Note that with this exposure, the statistical uncertainty is $\lesssim 1\%$ and somewhat smaller than the systematic uncertainties on the flux, detector, and backgrounds.
For $\mu$DAR, we assume a conservative $4\%$ detector and flux normalization systematics.

For solar neutrinos, the dominant backgrounds are beta and gamma activity from radon daughters in the uranium and thorium decay chains, and gamma cascades from the capture on argon of radiogenic neutrons produced by \((\alpha,n)\) reactions in the cavern rock; both mimic low-energy electron activity~\cite{Cuesta:2022mut, Capozzi:2018dat, MantheyCorchado:2024nao}.
We take the background spectra estimated for the DUNE far detector in Refs.~\cite{Cuesta:2022mut} (see also \cite{MantheyCorchado:2024nao}), rescaled to our exposure.
For the \(\mu\)DAR source, the dominant background comes from neutral-current interactions on argon induced by the \(\nu_e\) and \(\bar\nu_\mu\) fluxes from muon decay~\cite{COHERENT:2022nrm}.

The SN flux is what we aim to measure, and is therefore unspecified in \cref{tab:benchmarks}. 
We use a normalized pinched-thermal parametrization~\cite{Keil:2002in},
\begin{equation}
\label{eq:sn_flux}
\frac{1}{\Phi_0^\star}\frac{d\phi^\star}{dE_\nu}
=
\frac{\Phi^\star}{\Phi^\star_0}\,\mathcal N_\alpha
\left(\frac{E_\nu}{\langle E_\nu\rangle}\right)^\alpha
\exp\!\left[-(1+\alpha)\frac{E_\nu}{\langle E_\nu\rangle}\right],
\end{equation}
where $\mathcal N_\alpha =(1+\alpha)^{1+\alpha}/[\Gamma(1+\alpha)\langle E_\nu\rangle]$,  $\alpha$ is the pinching parameter,  $\langle E_\nu\rangle$ is the flux-averaged neutrino energy, and the nominal neutrino fluence at Earth is $\Phi^\star_0$.
For simplicity, we work with the flux integrated in time over the neutronization, accretion, and cooling phases, and we do not account for flavor oscillations~\footnote{To account for oscillations, the parametrization \eqref{eq:sn_flux} should be modified.}.

\textit{Spectral decomposition for the calibration sources.}---
While each calibration spectrum depends on the 200 weights through  \cref{eq:linear}, only certain linear combinations of weights that change the predicted spectrum appreciably are constrained by the data, combinations that barely affect it are not, but are also not needed to describe the measurement. 
The eigendecomposition \((R^\src)^T R^\src=U^\src\Lambda^\src(U^\src)^T\) separates the weight combinations by exactly this criterion.
Here, \(\Lambda^\src=\mathrm{diag}(\lambda^\src_1,\lambda^\src_2,\ldots)\) holds the eigenvalues in decreasing order and \(U^\src=(\mathbf u^\src_1,\bfm u^\src_2,\ldots)\) are the corresponding modes, which are linear combinations of different transitions.
A unit weight along the \emph{mode} \(\mathbf u^\src_\ell\) produces the spectrum \(R^\src\mathbf u^\src_\ell\), whose squared norm is the eigenvalue \(\lambda^\src_\ell\); the leading modes are therefore those with the largest imprint on the predicted spectrum.

For both calibration sources, the eigenvalues drop rapidly.
If we define the cumulative fraction
\(f^\src_m\equiv \sum_{\ell=1}^{m}\lambda^\src_\ell/\sum_{\ell=1}^{200}\lambda^\src_\ell\),
we find $(f^\odot_1,f^\odot_2)=(0.958,\,0.997)$ and $(f^\ominus_1,f^\ominus_3,f^\ominus_6)=(0.846,\,0.986,\,0.998)$.
This fast convergence follows from the smoothness of the calibration fluxes and detector responses, so only a few modes suffice to accurately describe the data.
We note that the backgrounds have a distinct shape from the leading modes of the signal in this description.

We define the \emph{mode weights} in the new basis as 
\begin{equation}
    w^\src_\ell\equiv (\mathbf{u}^\src_\ell)^T  \mathbf{w},
\end{equation}
and fit the data using
\begin{equation}
\label{eq:cal_trunc}
\mathbf n^\src_{(m_\src)}
=
\sum_{\ell=1}^{m_\src}(R^\src \mathbf u^\src_\ell)\,w^\src_\ell,
\end{equation}
to constrain the first \(m_\src\) mode coefficients.
Truncating at \((m_\odot,m_\ominus)=(2,6)\), the discarded modes change the predicted spectra by less than the statistical and systematic uncertainties of the calibration samples.

The calibration fits float the leading mode weights \(w^\src_\ell\) together with a signal-normalization nuisance absorbing the correlated overall flux uncertainty and, for each background template, a normalization and a linear tilt, all with Gaussian priors. 
The fits to calibration data yield best-fit mode weights
\begin{equation}
\label{eq:wcal}
\hatwcal = \widehat{\mathbf{w}}^\odot \oplus \widehat{\mathbf{w}}^\ominus =(\widehat{w}^\odot_1,\dots,\widehat w^\odot_{m_\odot},\widehat w^\ominus_1,\dots,\widehat w^\ominus_{m_\ominus})^T,
\end{equation}
and a covariance matrix $C_{\rm cal}=C^\odot\oplus C^\ominus$, block-diagonal across the two statistically independent sources, which are passed to the SN fit as a prior.
In the end, the leading $w^\src_\ell$ are constrained at the few-percent level; see End Matter for details.

\begin{figure}[th!]
    \centering
    \includegraphics[width=\linewidth]{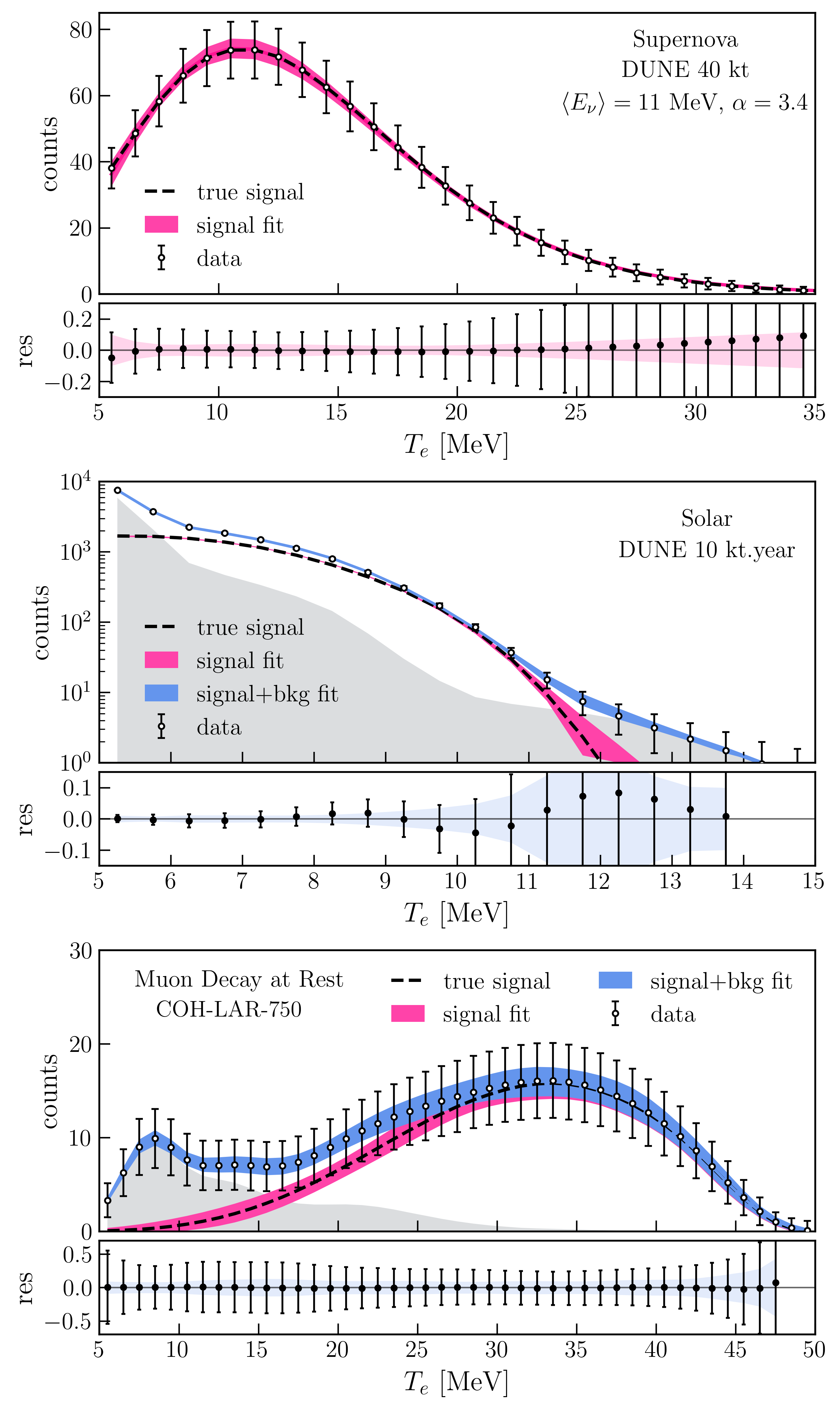}
    \caption{Spectrum-level fits to supernova (top), solar (middle), and $\mu$DAR (bottom) neutrinos using the data-driven method developed in this work. 
    The residual (res) subpanels show \((\mathrm{fit}-\mathrm{data})/\mathrm{data}\). Backgrounds are shown in gray.} 
    \label{fig:specfit}
\end{figure}

\textit{Cross section input for the SN measurement.}---The target of the SN fit is to measure the $\nu_e$ flux, here parametrized by $(\SNnorm, \alpha, \langle E_\nu\rangle)$; the cross section enters only through the prediction of the event spectrum. 
The SN burst is concentrated in a time window of $\lesssim 10$~s, so radioisotope backgrounds are negligible. 
The question is then to determine the combination of $\bfm{w}$ on which the SN spectrum depends. 
These are identified by the Fisher information matrix,
\begin{equation}
\label{eq:fisher}
    F^\star \equiv (R^\star)^T (C^\star_{T_e})^{-1} R^\star \, , 
\end{equation}
where $C^\star_{T_e}$ is the Poisson covariance of the predicted SN spectrum, dominated by the statistical uncertainty. 
The Fisher matrix can be diagonalized by $F^\star \bfm{u}_n^\star = \lambda_n^\star \bfm{u}_n^\star$ and its eigenvalues $\lambda^\star_n$ quantify the contribution and uncertainties of each mode to the SN spectrum.
Modes with large $\lambda^\star_n$ represent combinations that drive the predicted spectrum relative to its statistical noise, while modes with small $\lambda^\star_n$ neither affect the prediction nor are constrained by the data.
Because $R^\star$ and $C^\star_{T_e}$ depend on the SN flux parameters, the eigenmodes are recomputed at each point of the parameter scan. 
For the SN parameter ranges considered here, $m_\star = 3$ modes retain over $99\%$ of the Fisher information.

The solar, $\mu$DAR, and SN modes are directions in the same 200-dimensional weight space, with each set defined by the sensitivity of its own measurement. 
The calibration fits determine the components of $\bfm{w}$ along the solar and $\mu$DAR directions, while the SN fit needs its components along the SN directions.
How well the calibration constrains the SN fit is set by how much each SN direction overlaps with the subspace spanned by the $m_\odot$ and $m_\ominus$ calibration modes. 
Because the solar and $\mu$DAR energies bracket the SN spectrum, the leading SN modes lie close to the subspace spanned by the calibration modes.

To quantify this, we decompose each retained SN mode $\bfm{u}_n^\star$ into two pieces. 
First, a piece of $\bfm{u}_n^\star$ projected onto the calibration space, $U_{\rm cal}\bfm{q}_n$, where $U_{\rm cal} \equiv (\bfm u^\odot_1,\dots, \bfm u^\odot_{m_\odot}, \bfm u^\ominus_1,\dots, \bfm u^\ominus_{m_\ominus})$. 
Second, an orthogonal residual piece, $\bfm{r}_n$ which is the part that falls outside the calibration subspace.
Overall,
\begin{equation}
\label{eq:proj}
    \bfm{u}_n^\star = U_{\rm cal}\, \bfm{q}_n + \bfm{r}_n \, , \qquad p_n \equiv |U_{\rm cal}\, \bfm{q}_n|^2 \, .
\end{equation}
The projection fraction $p_n$ measures how well the calibration covers the $n$-th SN mode. 
If the latter lies entirely in the calibration subspace then $p_n=1$; the deviation from unity quantifies the component the calibration data cannot constrain, see End Matter~B for details.

As a benchmark for the SN $\nu_e$ flux, we choose true values $(\langle E_\nu\rangle, \alpha, \SNnorm) = (11~\text{MeV},\, 3.4,\, 1)$, with $\SNnorm$ rescaling the nominal time-integrated fluence $\Phi^\star_0 = 1.1\times 10^{11}$ neutrinos per cm$^2$~\cite{Capozzi:2018rzl}.
At the benchmark point, solar data alone cover the two leading SN modes at $p_{1,2}=(0.82,\,0.17)$ and $\mu$DAR alone at $p_{1,2}=(0.96,\,0.40)$, while their combination reaches $p_{1,2}=(0.998,\,0.989)$; the third mode is covered at $p_3=0.47$.

Re-expressed in the SN mode basis, the calibration fit becomes a Gaussian prior on the supernova mode weights $\bfm{w}^\star = (w^\star_1,\dots,w^\star_{m_\star})^T$ with central value and covariance
\begin{equation}
\label{eq:prior}
    \barwsn \equiv Q^T \hatwcal \, , \qquad
    C^\star_{\rm cal} \equiv Q^T C_{\rm cal}\, Q + \mathcal{C} \, ,
\end{equation}
where $Q = (\bfm q_1,\dots, \bfm q_{m_\star})$. The coverage term
\begin{equation}
\label{eq:cov}
    \mathcal{C}_{nm} = \frac{|\hatwcal|^2}{m_\odot+m_\ominus}\,
    \frac{\bfm{r}_n \cdot \bfm{r}_m}{\sqrt{p_n\, p_m}} \, ,
\end{equation}
accounts for the part of each SN mode that the calibration does not cover. 
Since $|\bfm r_n|^2 = 1-p_n$, the diagonal $\mathcal{C}_{nn}$ grows as $(1-p_n)/p_n$. 
When the calibration covers an SN direction well ($p_n \to 1$), $\mathcal{C}_{nm}$ vanishes for all $m$ and the mode inherits the calibration precision. 
As the projection fraction deviates from unity, the coverage degrades, the overall uncertainty on that mode enlarges, relying less on the calibration data.
To be conservative, whenever the Fisher information fraction of a SN mode $n$ is below 0.1\%, we enlarge the prior  on $w^\star_n$ by a factor 100.
In the region of interest considered here, the first two SN modes inherit the calibration precision and the third, due to its small Fisher information fraction, is left essentially unconstrained.

The final SN fit over binned data $\bfm{d}^\star$ is then
\begin{align}
\label{eq:sn_chi2}
    \chi^2 =\,
    & (\bfm{d^\star} - R^\star U^\star_{m_\star} \bfm{w}^\star)^T (C^\star_{T_e})^{-1}(\bfm{d^\star} - R^\star U^\star_{m_\star} \bfm{w}^\star) \nonumber\\
    & \quad+ (\bfm{w}^\star - \barwsn)^T (C^\star_{\rm cal})^{-1} (\bfm{w}^\star - \barwsn),
\end{align}
where $U^\star_{m_\star} \equiv (\bfm{u}^\star_1,\dots,\bfm{u}^\star_{m_\star})$, and the response $R^\star$, the eigenmodes, and the prior $(\barwsn, C^\star_{\rm cal})$ all depend on the SN parameters $(\SNnorm, \alpha, \langle E_\nu\rangle)$. 
Note that this $\chi^2$ can be minimized analytically.

\textit{Results.}---We apply the procedure above to mock data corresponding to a galactic SN at $10$~kpc detected at DUNE.
We then include calibration sources following the benchmark exposures in \cref{tab:benchmarks}: an early solar measurement at DUNE and a COH-Ar-750 $\mu$DAR sample.

\Cref{fig:specfit} shows the mock data, backgrounds, and resulting fits for the three sources.
We see that this description is not perfect, but remains well within the expected experimental uncertainties.
The SN spectrum, reconstructed using the projected calibration prior of \cref{eq:prior}, is in good agreement with the mock data and confirms that the prior captures the relevant cross section information for extracting the SN parameters.

\begin{figure}[t]
    \centering
    \includegraphics[width=\linewidth]{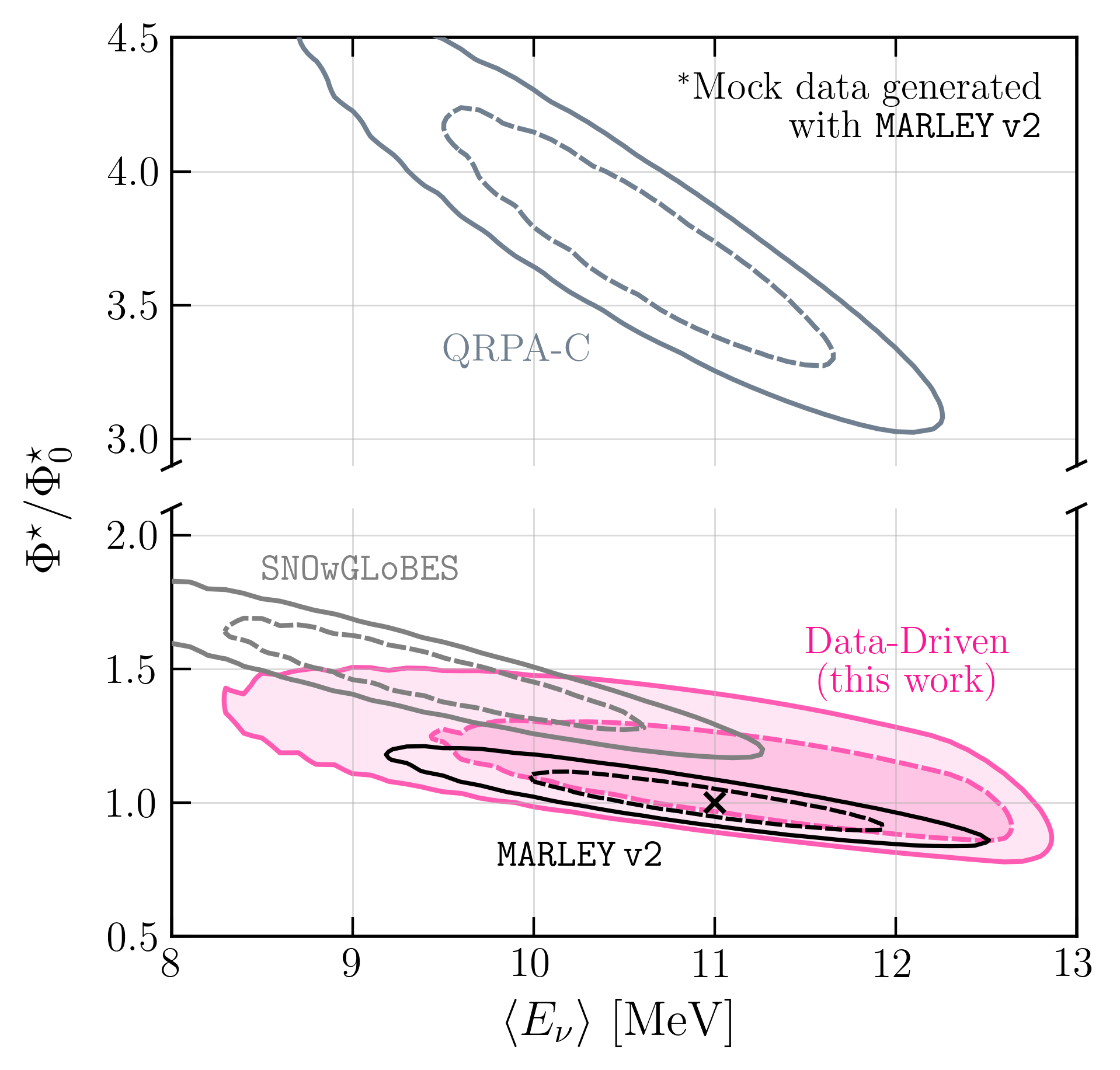}
    \caption{The $1\sigma$ (dashed) and $2\sigma$ (solid) contours in the $(\langle E_\nu\rangle, \SNnorm)$ plane after profiling over the pinching parameter $\alpha$. 
    The pink filled contour is the model-independent fit proposed in this work using the projected solar and $\mu$DAR priors of \cref{eq:prior}; 
    the other unfilled contours are the same fit but performed under different fixed cross section models.
    The mock data was generated with \MARLEYvtwo.
    }
    \label{fig:sn_contours}
\end{figure}

\Cref{fig:sn_contours} shows the preferred regions for the SN average energy $\langle E_\nu\rangle$ and fluence $\SNnorm$, profiled over the pinching parameter $\alpha$. 
We generate data assuming the \MARLEYvtwo~\cite{Gardiner:2026psy} cross section.
The fit with our data-driven approach is shown as the pink-shaded contours. 
For reference, we also show the fits obtained by fixing the cross section to the \MARLEYvtwo model, QRPA-C~\cite{Cheoun:2011zza}, and \texttt{SNOwGLoBES}~\cite{snowg} with no cross section uncertainties. 
The data-driven contour is competitive with the true-model fit and, more importantly, is centered on the truth parameters and with minimal bias.
The QRPA-C and \texttt{SNOwGLoBES} contours illustrate that a mismodeling of the cross section could substantially bias the extraction of the supernova spectrum.
This is to be expected given the large discrepancy among models~\cite{DUNE:2023rtr}.

\textit{Discussion.}---The procedure developed here achieves model-independent percent-level precision because extracting the SN spectral parameters requires knowledge of only a few linear combinations of the cross section weights, which are well covered by the calibration data.
In our method, the Fisher information content of the SN spectrum automatically identifies which combinations of nuclear matrix elements are relevant, being largely independent from model-specific input.
As such, it is designed to calibrate SN fits at the expense of not providing an explicit measurement of the cross section itself (although, see End Matter C).

To be useful, our method requires a few ingredients: 
i) the calibration flux must be well known, which is the case for solar and $\mu$DAR, 
ii) the template $\nu_e$CC differential cross section has enough support to include the true differential cross section, and 
iii) the SN $\nu_e$ flux is smooth enough that its Fisher information is accurately described by only a small number of spectral modes.
By construction, when the SN parameters stray outside the energy range bracketed by the calibrators, the coverage fractions $p_n$ drop, the prior inflates, and the fit to SN parameters no longer benefits from calibration priors.

Future improvements to our analysis include splitting the neutrino capture reaction into exclusive and semi-inclusive sample by tagging transitions to $\isotope[40]{K}^*$ states through de-excitation photons~\cite{DUNE:2020zfm, Gardiner:2021qfr, Gardiner:2026psy, Shi:2025rob}. 
Extending our method to general \emph{differential} cross sections should also reduce bias in the SN parameters.
Other improvements include accounting for oscillation effects in the SN spectrum, performing time-dependent SN fits, and implementing current and future data on specific transitions as priors for greater precision and robustness.

{\textbf{Notes added:} }As this paper was being completed, we learned of an independent study by Hajjar, Nairat, and Beacom~\cite{Hajjar:2026}.  
Both contributions were simultaneously submitted to arXiv.  Our two studies have connected but complementary approaches and results.

This work represents the views of the authors and not the DUNE Collaboration.

\begin{acknowledgments}
\textit{Acknowledgments.}---We thank Steven Gardiner for the fruitful discussions and feedback regarding \MARLEYvtwo.
We are especially grateful to Kevin Kelly and Stephan Meighen-Berger for a careful read of the manuscript.
We also thank John Beacom, Shirley Li, and Dan Pershey for discussions on this topic.
We acknowledge the use of Claude Code for coding and text review. N.M.\ is supported in part by US DOE Grant DE-SC0010813.
A.T.\ is supported in part by the DOE grant DE-SC0010143.
This work is partially supported by the University of Iowa's Year 2 P3 Strategic Initiatives Program through funding received for the project entitled ``High Impact Hiring Initiative (HIHI): A Program to Strategically Recruit and Retain Talented Faculty.''
This manuscript has been authored by Fermi Forward Discovery Group, LLC under Contract No. 89243024CSC000002 with the U.S. DOE, Office of Science, Office of High Energy Physics.
\end{acknowledgments}

\newpage

\onecolumngrid\medskip
\begin{center}\textbf{End Matter}\end{center}
\twocolumngrid
\appendix

\textit{A. Calibration likelihood.}---For each calibration source $\src\in\{\odot,\ominus\}$, the leading mode coefficients $\mathbf{w}^\src$  are obtained from
\begin{align}
\label{eq:em_cal_chi2}
\chi^2_\src(\mathbf{w}^\src)
&= \min_{\eta,\beta_q,\kappa_q}\Bigg[
\sum_i \frac{(d^\src_i - t^\src_i)^2}{d^\src_i}
+ \left(\frac{\eta-1}{\sigma_\eta}\right)^2 \nonumber\\
&\quad + \sum_q^{\rm bkgs} \left(\frac{\beta_q-1}{\sigma_{\beta,q}}\right)^2
+ \sum_q^{\rm bkgs} \left(\frac{\kappa_q}{\sigma_{\kappa,q}}\right)^2\Bigg],
\end{align}
with predicted spectrum
\begin{equation}
\label{eq:em_cal_pred}
  t^\src_i
  = \eta\,(R^\src U^\src_{m_\src} \mathbf{w}^\src)_i
+ \sum_q^{\rm bkgs} \beta_q\, b^\src_{qi}\,(1+\kappa_q\,\xi_i),
\end{equation}
where $U^s_{m_s} \equiv (\bfm{u}^s_1,\dots,\bfm{u}^s_{m_s})$, and the bin-centered tilt variable is $\xi_i = (T_{e,i}-\langle T_e\rangle)/(T_{e,\max}-T_{e,\min})$. 
Here, $d^\src_i$ is the mock data, $b^\src_{qi}$ is the $q$-th background template, and $\eta$ rescales the signal with prior width $\sigma_\eta$ which absorbs the correlated overall flux-normalization uncertainty. 
The rescaling and tilt of each background template is encoded in $\beta_q$ and $\kappa_q$, with corresponding uncertainties $\sigma_{\beta,q}$ and $\sigma_{\kappa,q}$. 

For solar neutrinos, the backgrounds $q$ consist of the radon-chain and neutron-capture backgrounds; while for $\mu$DAR, the dominant background is from neutral current interaction.
In this analysis we set $\sigma_\eta=4\%$ and $\sigma_{\beta,q}=\sigma_{\kappa,q}=10\%$ for all sources and background components. 
Although solar backgrounds largely exceed the signal in several regions,
see \cref{fig:specfit}, the leading solar eigenmodes are selected precisely because a background rescaling $\beta_q$ and tilt $\xi_i$ cannot reproduce their spectral shape. 
In fact, we have checked that this holds with if we rescale the backgrounds by 10 times.
Profiling Eq.~\eqref{eq:em_cal_chi2} over $(\eta,\beta_q,\kappa_q)$ at fixed $\mathbf{w}^\src$ leaves a $\chi^2$ that is quadratic in $\mathbf{w}^\src$, so the best-fit $\widehat{\mathbf{w}}^\src$ is obtained analytically and the post-profiling covariance is the inverse Hessian
\begin{equation}
\label{eq:em_Cw}
(C^\src)^{-1}_{ij} = \tfrac{1}{2}\,\frac{\partial^2 \hat{\chi}^2_\src}{\partial w^\src_i\,\partial w^\src_j}\bigg|_{\widehat{\mathbf{w}}^\src},
\qquad
\hat\chi^2_\src(\mathbf{w}^\src) \equiv \min_{\eta,\beta_q,\kappa_q} \chi^2_\src.
\end{equation} 
The calibration prior in \cref{eq:prior} of the main text uses the best-fit mode coefficients of both sources, concatenated into
\begin{equation}
\hatwcal = (\widehat w^\odot_1,\dots,\widehat w^\odot_{m_\odot},\widehat w^\ominus_1,\dots,\widehat w^\ominus_{m_\ominus})^T,
\end{equation}
with covariance
\begin{equation}
\label{eq:em_Ccal}
C_{\rm cal} = C^\odot \oplus C^\ominus,
\end{equation}
block-diagonal because the two calibration datasets are statistically independent. The mode counts $(m_\odot,m_\ominus)$ entering Eq.~\eqref{eq:em_Ccal} are chosen as the largest values for which all included modes remain well constrained by their respective calibration data; modes with small eigenvalues that would inject large bias are excluded.

\begin{figure}[t]
  \centering
  \includegraphics[width=\linewidth]{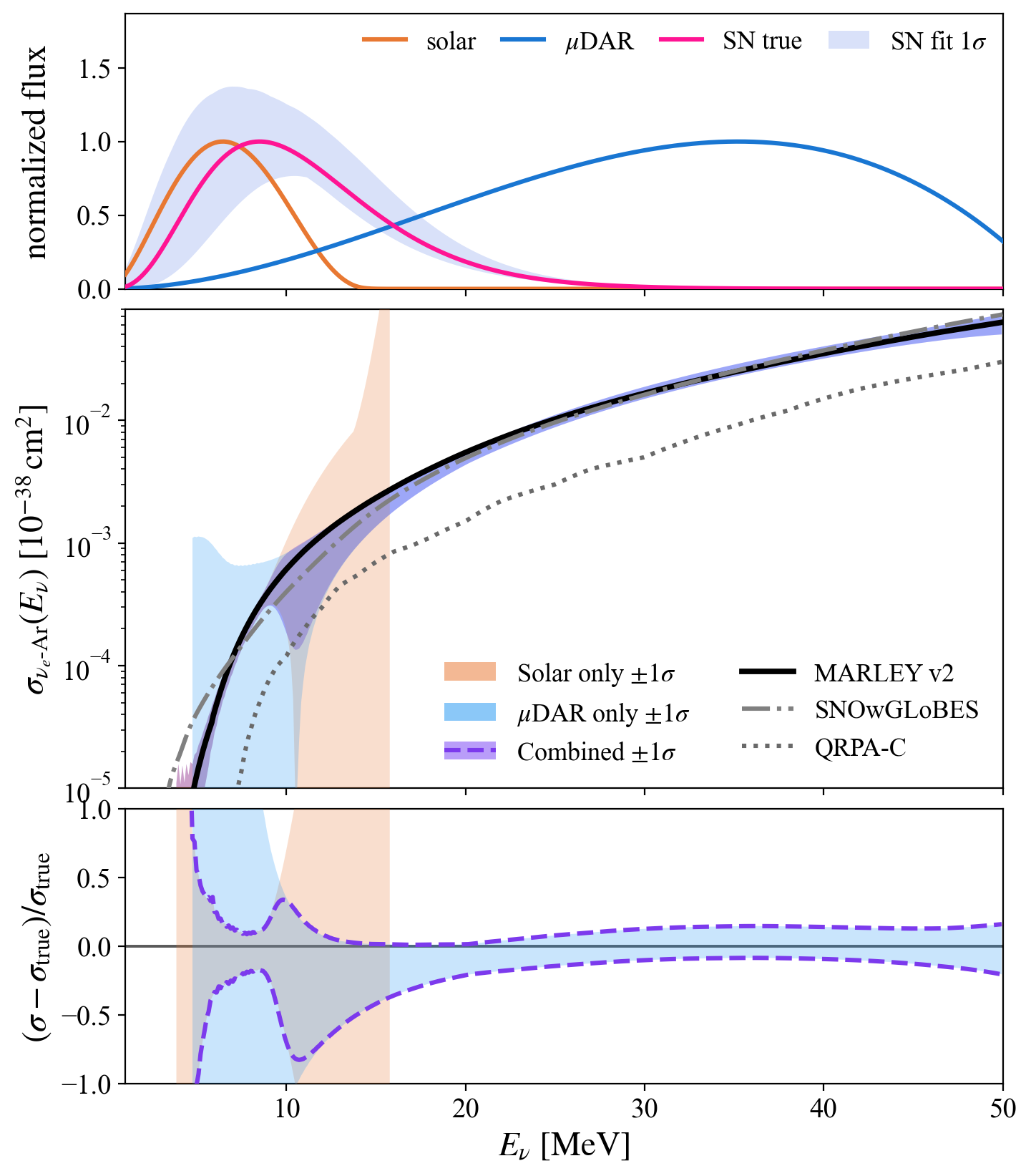}
  \caption{Calibration-derived $1\sigma$ constraints on $\sigma_{\nu_e\text{-Ar}}(E_\nu)$ for mock data generated with the \MARLEYvtwo cross section (solid black). The orange band (solar only) covers the $^8$B flux--response window; the blue band ($\mu$DAR only) covers the higher-energy window; the purple band (combined) uses inverse-variance weighting. Reference theory curves for QRPA-C~\cite{Cheoun:2011zza} and \texttt{SNOwGLoBES}~\cite{snowg} are shown for comparison.}
  \label{fig:xsec_bands}
\end{figure}

\textit{B. Projection onto the SN Fisher eigenmodes.}---The retained calibration eigenmodes form
\begin{equation}
\label{eq:em_Ucal}
U_{\rm cal} \equiv (u^\odot_1,\dots,u^\odot_{m_\odot}, u^\ominus_1,\dots,u^\ominus_{m_\ominus}),
\end{equation}
a $200\times(m_\odot+m_\ominus)$ matrix. 
Within each source the columns are orthonormal eigenvectors of $(R^\src)^T R^\src$, but across sources they are generically non-orthogonal, e.g. $u^\odot_1\cdot u^\ominus_2\neq 0$. 
Because $U_{\rm cal}$ does not have orthonormal columns, the orthogonal projector onto $\mathrm{span}(U_{\rm cal})$ is $P_{\rm cal}=U_{\rm cal}\,G_{\rm cal}^{-1}\,U_{\rm cal}^T$ with $G_{\rm cal}\equiv U_{\rm cal}^T U_{\rm cal}$ the Gram matrix; the factor $G_{\rm cal}^{-1}$ corrects for the non-orthogonality (compensating for the possible double coverage) between the solar and $\mu$DAR blocks.
For each retained SN Fisher eigenmode we define
\begin{equation}
\label{eq:em_qn}
\bfm{q}_n \equiv G_{\rm cal}^{-1} U_{\rm cal}^T \bfm{u}_n^\star,
\qquad
\bfm{r}_n \equiv \bfm{u}_n^\star - U_{\rm cal}\,\bfm{q}_n,
\end{equation}
so that the residual $r_n$ is orthogonal to $\mathrm{span}(U_{\rm cal})$ and the projection fraction in \cref{eq:proj} is
\begin{equation}
\label{eq:em_pn}
p_n = |U_{\rm cal}\,\bfm{q}_n|^2 = (\bfm{u}_n^\star)^T U_{\rm cal}\,G_{\rm cal}^{-1}\,U_{\rm cal}^T \bfm{u}_n^\star.
\end{equation}

\textit{C. Cross section constraint from the calibration fit.}---The calibration fit described in Section~A determines $\widehat w_{\rm cal}$ and its covariance $C_{\rm cal}$ from Eq.~\eqref{eq:em_Ccal}. Back-projecting to the original $\Delta E_k$ basis via
\begin{equation}
  \hatw = U_{\rm cal}\,\hatwcal,
  \qquad
  C_w = U_{\rm cal}\,C_{\rm cal}\,U_{\rm cal}^T,
\end{equation}
and substituting into \cref{eq:xsec} of the main text yields the implied total cross section.
Defining the kinematic factor
\begin{equation}
\label{eq:em_gk}
  g_k(E_\nu) \equiv
  \int dT_e\,
  \frac{E_e p_e}{(1~\text{MeV})^2}
  \,\delta(E_e - E_\nu + \Delta E_k),
\end{equation}
the implied total cross section is
\begin{equation}
\label{eq:em_sigma_proj}
  \hat\sigma(E_\nu) = \sigma_0 \sum_k \hat{w}_k\, g_k(E_\nu),
\end{equation}
with propagated $1\sigma$ uncertainty
\begin{equation}
\label{eq:em_sigma_var}
  \mathrm{Var}[\hat\sigma(E_\nu)]
  = \sigma_0^2\,g(E_\nu)^T C_w\, g(E_\nu) \, .
\end{equation}
We use the full $\Delta E_k$ basis ($m_\odot = m_\ominus = 200
$), so no directions in $w$-space are discarded. Directions with negligible Fisher information, where the data cannot constrain $w_k$, are handled by a zero-mean box prior $w_k \sim \mathrm{Uniform}[-w_{\rm b},\,w_{\rm b}]$ with $w_{\rm b}=20\max(|\widehat w^\odot_1|,|\widehat w^\ominus_1|)$. 
This widens the band at energies where the flux provides little leverage, rather than artificially tightening it by discarding those directions. 
Energies outside each source's flux support ($\leq 10^{-4}\%$ of the total flux) are masked entirely and no band is drawn there.

\Cref{fig:xsec_bands} shows the resulting $1\sigma$ bands for the solar-only, $\mu$DAR-only, and combined fits. 
The combined $1\sigma$ fractional uncertainty is shown in the lower panel 
and listed in \cref{tab:xsec_bands}. 
In the energy region where the cross section is determined by the solar data, the uncertainty reaches a minimum of $\sim 8\%$ near the peak of the solar flux. 
The $\mu$DAR energy spectrum constrains the cross section in a broader region of energies above $15$~MeV. 
In the intermediate region, which draws partial support from both spectra, the cross section is more weakly constrained, but better than either source could achieve alone. 
We can observe significant differences between QRPA-C predictions, \texttt{SNOwGLoBES} evaluations and \MARLEYvtwo. 
A calibration from solar and $\mu$DAR data here would therefore have the power to discriminate among nuclear-theory predictions, independently of the SN analysis.
A full cross section extraction and model comparison is left for  future work.

\begin{table}[t]
  \caption{Fractional $1\sigma$ uncertainty on the calibration-projected $\sigma_{\nu_e\text{-Ar}}(E_\nu)$.}
  \label{tab:xsec_bands}
  \begin{ruledtabular}
    \begin{tabular}{cccc}
      $E_\nu$ [MeV] & Solar [\%] & $\mu$DAR [\%] & Combined [\%] \\
      \hline
       8 &   8 &  599 &   8 \\
      10 &  44 &  160 &  43 \\
      12 &  159 &  65 &  61 \\
      15 &  1789 &  26 &  26 \\
      20 & --- &  10 &  10 \\
      34 & --- &  7 &  7 \\
    \end{tabular}
  \end{ruledtabular}
\end{table}

\bibliography{main}

\end{document}